\documentclass[pra, twocolumn]{revtex4}
\usepackage{amstext,amsmath,amssymb,amsfonts}
\usepackage{bm,bbm,euscript}
\usepackage{epsfig}
\usepackage{verbatim}
\usepackage{latexsym}
\usepackage{booktabs}

\def\1{{{\mathbbm 1}}}
\def\6{\langle}
\def\9{\rangle}

\def\tr{{\rm tr}\,}

\def\aP{\mathsf{P}}

\def\half{\mbox{$1\over2$}}
\def\quar{\mbox{$1\over4$}}

\def\be{\begin{equation}}
\def\ee{\end{equation}}

\def\cH{{\cal H}}

\begin{document}
\title{Quantum discord and local demons\footnote{Published as ``Quantum discord, local operations and Maxwell's demons"}}
\author{Aharon Brodutch}
\email{aharon.brodutch@science.mq.edu.au}
\author{Daniel R. Terno}
\email{daniel.terno@mq.edu.au}
\affiliation{Department of Physics, Faculty of Science, Macquarie University, NSW 2109, Australia}


\begin{abstract}
Quantum discord was proposed as a measure of the quantumness of correlations. There are at least three different discordlike quantities, two of which determine the difference between the efficiencies of a Szilard's engine under different sets of restrictions. The three discord measures vanish simultaneously.  We introduce an easy way to test for zero discord, relate it to the Cerf-Adami conditional entropy and show that there is no  simple  relation between the discord and local distinguishability.
\end{abstract}

\maketitle

\section{Introduction}
We learn about the external world from the correlations between our measuring devices and the systems we study. The information flow is studied in (classical) information theory  through the analysis of correlations \cite{tc}. Their build-up and propagation are at the core of measurement theory \cite{mes, peres}, while the research of their behavior with respect to  spacetime locality led to the identification of quantum entanglement and was one of the precursors of quantum information theory \cite{peres, nc}. All correlations --- classical and quantum --- are important in statistical thermodynamics \cite{qter}, which in turn influences the entanglement theory \cite{horbig}. Initial correlations modify the dynamics of open systems  \cite{dhk,discp}, and a significant part of quantum information processing is devoted to preservation and manipulation of useful correlations, while  trying to mitigate the effects of the unwanted ones.

A boundary between  quantum and classical correlations  is  sharp in pure states, which are  either simply separable product states  or entangled. It becomes less clear in mixed states, particularly for  systems larger than a pair of qubits. We investigate this boundary through a characteristic of quantum discord \cite{oz01,hv01,dat} (more precisely, we review and define three similar,  but subtly different discord measures).

First , we recall some basic definitions \cite{tc} and set the notation. The (Shannon) entropy of a classical discrete  probability distribution $p(a)\equiv p_a$ over a random variable $A$ is defined by
\be
H(A)=-\sum_a p_a\log p_a,
\ee
where the base of the logarithm is determined by conventions. Entropy of the joint probability distribution $p(a,b)$ over $AB$, $H(AB)$  is defined analogously.  The Bayes theorem relates it to the conditional probabilities,
\be
p(a,b)=p(a|b)p(b)=p(b|a)p(a),
\ee
where $p(a|b)$ is a conditional probability of $A=a$ given that $B=b$. The conditional entropy of $A$
\be
H(A|B)=\sum_b p_b H(A|b)=-\sum_{a,b}p(a,b)\log p(a|b)
\ee
is a weighted average of the entropies of $A$ given a particular outcome of $B$.

Correlations between  two probability distributions are measured by the symmetric mutual information. It has two equivalent expressions,
\be
I(A:B)=H(A)+H(B)-H(A,B), \label{mi}
\ee
and
\be
J(A:B)=H(A)-H(A|B)=H(B)-H(B|A). \label{mj}
\ee

Quantum-mechanical (von Neumann) entropy \cite{peres} is defined as
\be
S(\rho)=-\tr\rho\log\rho.
\ee
It minimizes the Shannon entropy of  probability distributions that result from rank-1 positive operator-valued measures (POVMs) that are applied to the state $\rho$ on the Hilbert space $\cH_A$. The minimum is actually reached on a probability distribution $A$ that results from a projective measurement $\Pi=\{\Pi_a, a=1,\ldots d\}$, $\sum_a\Pi_a=\1$, $\Pi_a\Pi_b=\delta_{ab}\Pi_a$, which is constructed from the eigenprojectors of $\rho$,
\be
S(\rho)=\min_\Pi H(A_\rho^\Pi) \label{minent},
\ee
that is,
\be
S(\rho)=H(A_\rho^{{\Pi^*}}), \qquad \rho=\sum_a p_a{\Pi}_a^*, \quad p_a\geq 0, \sum_a p_a=1.
\ee
The expression $A_\rho^\Lambda$ stands for a classical probability distribution (of a measured parameter $A$) that is obtained from the state $\rho$ under the POVM $\Lambda$.

Quantum channels are abstracted as maps (typically completely positive ones) from the initial states $\rho_A$ to the final states $\rho_X$, where the space $\cH_X$ may be either the same space $\cH_A$ or different one \cite{nc}. Any orthonormal basis $\{|a\9\}$ of $\cH_A$ defines a dephasing channel $\aP$,
\be
\rho\mapsto \rho'=\aP(\rho)=\sum_a\Pi_a\rho\Pi_a, \qquad \Pi_a:=|a\9\!\6a|.
\ee
Hence, taking the weighted average over the outcomes of a measurement $\Pi=\{\Pi_a\}_{a=1,\ldots d_A}$  is equivalent to sending the initial  state through a dephasing channel with a superoperator $\aP=\aP_\Pi$.

The information function that expresses   knowledge \cite{horbig} about a system of dimension $d$ in the state $\rho$ is a variety of the negentropy,
\be
K(\rho)=\log d-S(\rho).
\ee
It has two related operational interpretations.   A Maxwell's demon can  draw $K(\rho)$ work   from a single heat bath using a Szilard engine \cite{md2} by using pure states. We will discuss the demons in Sec.~III.

 On the other hand, $K(\rho)$ determines a conversion rate between pure and mixed states. By allowing arbitrary unitary operations, by adding of maximally mixed ancillas  and by taking partial traces (and thinking in terms of qubits, $\log 2=1$), it is possible to perform two tasks with asymptotically perfect fidelity. First, given $n$ copies of the state $\rho$ one can obtain $nK(\rho)$ qubits in a pre-determined pure state. This is done by performing the usual quantum data compression, but instead of discarding the then-redundant pure qubits, it is the signal that is discarded \cite{horbig}. Second, by taking $nK(\rho)$ pure qubits and by diluting them with ancillas in the maximally mixed state, one produces $n$ copies of $\rho$ \cite{hho}.

Quantum discord stems from the fact that the classical mutual information can be extended to  quantum states in two inequivalent ways, following either Eq.~\eqref{mi} or Eq.~\eqref{mj}. In Sec. II, we introduce the three discord measures and discuss some of their properties. Their role in the efficiency of different Szilard's engines is explored in Sec.~III, and, in Sec.~IV,  we discuss their relationship to  the local distinguishability of orthogonal states.

\section{Quantum discord}
The first expression for mutual information has an obvious quantum generalization,
\be
I(\rho_{AB}):=S(\rho_A)+S(\rho_B)-S(\rho_{AB}),
\ee
and represents the total amount of quantum and classical correlations \cite{horbig,gpw,vl10}.

\subsection{Conditional ``state" definition}

To obtain a quantum version of $J(A:B)$, it is necessary to determine a conditional state of the subsystem $B$. If the objective is to preserve the equivalence of two definitions in the quantum domain, $I(\rho_{AB})\equiv J(\rho_{AB})$, then the conditional entropy can be introduced \cite{cerfadami} as
\be
S(\rho_{B|A})=-\tr\rho_{AB}\log\rho_{B|A},
\ee
The positive operator $\rho_{B|A}$ is defined through
\begin{align}
\rho_{B|A}&:=\lim_{n\rightarrow\infty}(\rho_{AB}^{1/n}(\rho_A\otimes\1_B)^{-1/n})\nonumber\\&=\exp(-\log\rho_A\otimes\1_B+\log\rho_{AB}),
\end{align}
where the inverse of $\rho_A$ is defined on its support. It does not usually have a unit trace, and when $\rho_A\otimes\1_B$ commutes with $\rho_{AB}$ it reduces to
\be
\rho_{B|A}=\rho_{AB}(\rho_A\otimes\1_B)^{-1}.
\ee
We will return to this quantity at the end of this section.

\subsection{Three versions of the discord}
Given a complete projective measurement $\Pi$ on $A$, a quantum definition of $J$ follows its interpretation as the information gained about the system $B$ from the measurement on $A$ \cite{oz01},
\be
J^{\Pi^A}(\rho_{AB}):=S(\rho_B)-S(\rho_B|\Pi^A),
\ee
where the conditional entropy is now given by
\be
S(\rho_B|\Pi^A):=\sum_a p_a S(\rho_{B|\Pi_a}).
\ee
The post-measurement state of $B$ that  corresponds to the outcome $A=a$  is
\be
\rho_{B|\Pi_a}= (\Pi_a\otimes\1_B \rho_{AB}\Pi_a\otimes\1_B)/p_a, \qquad p_a=\tr\rho_A\Pi_a. \label{def-b}
\ee
The state of $B$ remains unchanged,
\be
\rho_B=\tr_A\rho_{AB}=\sum_a p_a\rho_{B|\Pi_a}.
\ee

Unlike their classical counterparts, the quantum expressions are generally inequivalent and $I(\rho_{AB})\geq J^{\Pi^A}(\rho_{AB})$ \cite{oz01,hv01,dat}. The \textit{quantum discord} as defined in \cite{oz01}  is the difference between these two quantities,
\be
D_1^{\Pi^A}(\rho_{AB}):=S(\rho_A)+S(\rho_B|\Pi^A)-S(\rho_{AB}).
\ee
Its dependence on the measurement procedure is removed by minimizing the result over all possible sets of $\Pi$,
\be
D_1^A(\rho_{AB}):=\min_{\Pi^A} D_1^{\Pi^A}(\rho_{AB}).
\ee
Similarly,
\be
J_1^A(\rho_{AB}):=\max_{\Pi^A}J^{\Pi^A}(\rho_{AB}).
\ee
This definition of the discord has its origins in the studies of the measurement procedure and pointer bases,  thus projective measurements are natural in this context. It is possible to define the discord when the difference is minimized over all possible  POVM $\Lambda^A$ \cite{hv01}. However, unless stated otherwise we restrict ourselves to the projective measurements of rank 1.

An explicit form of a post-measurement state will be useful in the following text. We denote this state as $\rho'_{X}\equiv\rho^{\Pi^A}_X$, where the subscript $X$ stands for $A$, $B$, or $AB$, and  use the former expression if it does not lead to  confusion. After a projective measurement $\Pi^A$, the state of the system becomes
\be
\rho_{AB}'=\sum_a p_a\Pi_a\otimes\rho^a_B,  \label{postmes}
\ee
where $p_a$ and $\rho_B^a\equiv\rho_{B|\Pi_a}$ are given by Eq.~\eqref{def-b}, and the states of the subsystems are
\be
\rho'_A=\sum_a p_a\Pi_a, \qquad \rho_B=\rho_B'=\sum_a p_a\rho^a_B,
\ee
respectively.

The discord of the state $\rho_{AB}$ is zero if and only if it is a mixture of products of arbitrary states of $B$ and projectors on $A$ \cite{oz01},
\be
\rho_{AB}=\sum_a p_a\Pi_a\otimes\rho^a_B, \qquad  p_a\geq 0, \quad \sum_a p_a=1. \label{zerodisc}
\ee

By using this decomposition and properties of the entropy of block-diagonal matrices \cite{wehrl} we can identify
\be
J^{\Pi^A}(\rho_{AB})\equiv I(\rho_{AB}'),
\label{mutinfo}
\ee
because $S(\rho_A')=H(A_\rho^\Pi)$ and
\be
S(\rho_{AB}')=H(A_\rho^\Pi)+S(\rho_B|\Pi^A). \label{ent-post}
\ee

The discord is not a symmetric quantity: It is possible to have states  with $D_1^A(\rho_{AB})\neq D_1^B(\rho_{AB})$.  A subclass of separable states that satisfy $D_1^A=D_1^B=0$ is of the form
\be
\rho_{AB}^c=\sum_{ab} w_{ab}\Pi_a^A\otimes P^B_b,
\ee
where $P^B$ is  a set of projectors on $\cH_B$, and consists of  classically correlated states in the sense of \cite{ohhh89-18}.

Another possibility  is to set
\begin{align}
{J}_2^{\Pi^A}&:=S(\rho_A)+S(\rho_B)- [H(A_\rho^\Pi)+S(\rho_B|\Pi^A)]\nonumber\\&=S(\rho_A)+S(\rho_B)-S(\rho_{AB}'),
\end{align}
arriving  at the quantum discord as defined in \cite{z03}
\be
D_2^A(\rho_{AB}):=\min_\Pi[H(A_\rho^\Pi)+S(\rho_B|\Pi^A)]-S(\rho_{AB}), \label{def_d2}
\ee
where the quantity to be optimized is a sum of post-measurement entropies of $A$ and $B$.
By using Eq.~\eqref{minent}, we see that $D_1\leq D_2$. It is also easy to see that $D_1=0 \Leftrightarrow D_2=0$. By using Eqs.~\eqref{postmes} and \eqref{ent-post}, we obtain a different expression for $D_2$:
\be
D_2^{\Pi^A}(\rho_{AB})=S(\rho_{AB}^{\Pi^A})-S(\rho_{AB}). \label{ent_inc2}
\ee

Since the definition of the discord(s) involves  optimization, the analytic expressions are known only in some particular case \cite{oz01,dat,luo08}. Moreover, typically it is important to know whether the discord is zero or not, while the numerical value itself is less significant.

It follows from Eq. \eqref{zerodisc}  that if the spectrum of a reduced state $\rho_A=\sum_a p_a\Pi_a$ is non-degenerate, then its  eigenbasis gives a unique family of projectors $\Pi$ that results in the zero discord for $\rho_{AB}$. Hence, a recipe for testing states for zero discord and for finding the optimal basis is to trace out a subsystem that is left alone ($B$),  to diagonalize $\rho_A$ and to calculate the discord in the resulting eigenbasis.

If the state $\rho_A$ is degenerate, a full diagonalization should be used. For the state with the form of Eq.~\eqref{zerodisc},  each of the reduced states $\rho^a_B$ can be diagonalized as
\be
\rho^a_B=\sum_b r^a_b P^b_a, \qquad P^b_a P^{b'}_a=\delta^{bb'}P^b_a.
\ee
The eigendecomposition of the state $\rho_{AB}$ then easily follows. Writing it as
\be
\rho_{AB}=\sum_{a,b}w_a r^a_b \Pi_a\otimes P^b_a, \label{two0}
\ee
it is immediate to see that its eigenprojectors are given by $\Pi_a\otimes P^b_a$. Hence, if $\rho_B$  has a degenerate spectrum, but $\rho_{AB}$ has not, the structure of its eigenvectors reveals if it is of a zero or nonzero discord. Hence we established

\textbf{Property 1.}
The eigenvectors of a zero discord state $D_1^A(\rho_{AB})=0$ satisfy
\be
\rho_{AB}|ab\9=r_{ab}|ab\9, \quad\Rightarrow\quad |ab\9\6ab|=\Pi_a\otimes P^b_a.
\ee
$~$ \hfill $\Box$

This consideration leads to the simplest necessary condition for zero discord (first noticed in \cite{fc09}):

\textbf{Property 2.} If $D_1^A(\rho_{AB})=0$, then
\be
[\rho_A\otimes\1_B,\rho_{AB}]=0.
\ee
Hence a non-zero commutator implies $D_1^A(\rho_{AB})>0$. \hfill $\Box$

Naturally, if the state has  zero discord, and the eigenbasis is only partially degenerated, we can use it to reduce the optimization space. On the other hand, the diagonalizing basis  $\Pi_*$ is not necessarily the optimal basis $\hat{\Pi}$ or  $\check{\Pi}$ that enters the definition of $D_1$ or $D_2$, respectively.  Consider, for example, a two-qubit state
\be
\rho_{AB}=\quar(\1_{AB}+b \sigma^z_{A}\otimes\1_B+c\sigma^x_{A}\otimes\sigma^x_{B}),\label{ex_disc}
\ee
where $\sigma^a_X$ are Pauli matrices on the relevant spaces, $X=A,B$, and the constants $b$ and $c$ are restricted only by the requirements that $\rho_{AB}$ is a valid density matrix. For this state $\rho_B=\1/2$ and $\rho_A=\mathrm{diag}(1+b,1-b)/2$. After the measurement in the  diagonalizing basis $\Pi^z=((\1+\sigma^z)/2,(1-\sigma^z)/2)$ the conditional state of $B$ becomes
\be
\rho_{B|\Pi^z_\pm}=\1/2,
\ee
and the conditional entropy is maximal, $S(\rho_B|\Pi^z)=\log 2$.

On the other hand, in the basis $\Pi^x=((\1+\sigma^x)/2,(1-\sigma^x)/2)$ the probabilities of the outcomes are equal, $p_+=p_-=1/2$, but the post-measurement states of $B$ are different from the maximally mixed one,
\be
\rho_{B|\Pi^x_\pm}=\half(\1\pm c \sigma^x),
\ee
so the entropy $S(\rho_B|\Pi^z)\geq S(\rho_B|\Pi^x)$.

This discrepancy motivates us to  define  a new version of the discord, which is useful if the eigenvectors of $\rho_A$ are not degenerate,
\be
D_3^{A}(\rho_{AB}):=S(\rho_A)-S(\rho_{AB})+S(\rho_B|\Pi^A_*),
\ee
where $\Pi^A_*$ is the set of eigenprojectors of $\rho_{A}$. Otherwise it can be introduced using the continuity of entropy in finite-dimensional systems \cite{wehrl}.
By applying Eq.~\eqref{minent} to the subsystem $A$ we find that  $D_3$ simultaneously holds the analog of Eq.~\eqref{mutinfo},
\be
J_3(\rho_{AB})\equiv I(\rho_{AB}^{\Pi_*^A}),
\ee
and of Eq.~\eqref{ent_inc2},
\be
D_3(\rho_{AB})\equiv S(\rho_{AB}^{\Pi^A_*})-S(\rho_{AB}).
\ee

We also arrive to the following ordering of the discord measures:
\be
D_1^A\leq D_2^A\leq D_3^A \label{duneq}.
\ee
There are several important cases when the  measures of discord coincide. Most importantly, they vanish simultaneously:

\textbf{Property 3.} $D_1=0\Leftrightarrow D_2=0\Leftrightarrow D_3=0$.

\noindent The proof follows from Eqs.\eqref{zerodisc} and \eqref{duneq}. \hfill $\Box$.

 On pure states
 the discord is equal to the degree of entanglement,
\be
D_i^A(\phi_{AB})=S(\phi_A)=E(\phi_{AB}), \qquad i=1,2,3.
\ee
Discord is also independent of the basis of measurement if the state is invariant under local rotations \cite{oz01}. Finally,
if $A$ is in a maximally mixed state, then $D_1^A=D_2^A$.

These coincidences make it interesting to check when the discords $D_1$ and $D_2$ are different. By returning to the measurement-dependent versions of the discords, we see that
\be
D_1^{\Pi^A}(\rho_{AB})=D_2^{\Pi^A}(\rho_{AB})-[H(A^{\Pi}_\rho)-S(\rho_A)].
\ee
Assume that  $D_2^{\Pi^A}(\rho_{AB})$ reaches the minimum on the set of projectors $\check{\Pi}$, which are not the eigenprojectors of $\rho_A$. In this case $H(A^{\check{\Pi}}_\rho)-S(\rho_A)>0$, so we can conclude that the strict inequality $D_1^A<D_2^A$ holds, because
\begin{align}
D_1^A(\rho_{AB})&\leq D_1^{\check{\Pi}}(\rho_{AB}) \\
=&D_2^A(\rho_{AB})-[H(A^{\check{\Pi}}_\rho)-S(\rho_A)]<D_2^A(\rho_{AB})\nonumber.
\end{align}
For example, the state of Eq.~\eqref{ex_disc}, with $b=c=\half$, satisfies $D_1^A\approx 0.05$, $D_2^A\approx 0.20$ and $D_3^A\approx 0.21$.

\subsection{Relations with other quantities}

Quantum discord $D_1^{\Lambda^A}(\rho_{AB})$ is a concave function over the set of all POVMs $\Lambda^A$ \cite{dat} and the minimum is reached on a rank-1 POVM that consists of linearly independent operators \cite{povm}. The easiest way to obtain this result is to note that the set of all POVMs is convex, and the minimum of a concave function over a convex set is obtained on its boundary.

The states of zero discord are nowhere dense in the set of all states \cite{fc09}. Nevertheless, it is obvious that even double-zero discord states of Eq.~\eqref{two0} convexly span the set of all states.

The operator $\rho_{B|A}$ has a closed form on the states of zero discord. Property 1 allows to write it as
\begin{align}
\rho_{B|A}&=\left(\sum_a w_a\Pi_a\otimes\rho_B^a\right)\left(\sum_b \frac{1}{w_b}\Pi_b\otimes\1_B\right)\nonumber\\&=\sum_a\Pi_a\otimes\rho_B^a,
\end{align}
which indeed results in the conditional entropy
\be
S(\rho_{B|A})=\sum_a w_a S(\rho_B^a)=S(\rho_{B}|\Pi^A).
\ee
In general it follows from the definitions that
\be
S(\rho_{B|A})=S(\rho_{B}|\Pi^A)-D^{\Pi^A}_1(\rho_{AB}),
\ee
in any basis, not only in the optimal one.

In the paradigm of closed local operations \cite{horbig}, Alice and Bob are allowed to perform arbitrarily local unitaties and projective measurements, and Alice can send her system to Bob via a dephasing channel. In the one-way version only a single  use of the channel is allowed. Since at the end of the operation both system are accessible to Bob, the discord $D_2^A$ was identified with one-way quantum deficit \cite{horbig},
\be
\Delta^\rightarrow(\rho_{AB})=\min_{\Pi^A} S(\rho_{AB}^{\Pi^A})-S(\rho_{AB})=D_2^A(\rho_{AB}).
\ee
Operationally, it expresses the fact that a naive one-way purification strategy consists of  Alice performing the measurement $\check{\Pi}$ and announcing her results to Bob (or, equivalently, sending her part of the state individually through the channel $\aP_{\check{\Pi}}$). In this case the purification rate is given by $K_2(\rho_{AB})=\log d_Ad_B- S(\rho_{AB}^{\check{\Pi}^A})$.

However, it is the discord $D_1^A$ that gives the optimal efficiency of a purification in this context. If Alice is allowed to borrow pure states (that are returned at the completion of the protocol) and use block encoding  prior to (individually) sending her particles and Bob, then \cite{dev05} the optimal rate is
\be
K^\rightarrow(\rho_{AB})=\log d_Ad_B+I(\rho_{AB}^{\hat{\Pi}})-S(\rho_A)-S(\rho_B),
\ee
so using Eq.~\eqref{mutinfo} we see that $K^\rightarrow(\rho_{AB})=\log d_Ad_B+J_1^A(\rho_{AB})-S(\rho_A)-S(\rho_B)$,
or
\be
K(\rho_{AB})-K^\rightarrow(\rho_{AB})=D_1^A(\rho_{AB}).
\ee
Since the additivity of $J_1^A$ was shown to be equivalent to several other additivity problems \cite{dev05}, including the Holevo capacity of quantum channels, and the latter was finally disproved \cite{hast09},  block processing will improve the efficiency for entangled $\rho_{AB}$.

A symmetrized version of $J_3$ involves (projective) measurements on both sides \cite{luodatta} and is given by $J_3^\mathrm{{sym}}(\rho_{AB})=I\bigl(\rho_{AB}^{\Pi^A_*\otimes\Pi^B_*}\bigr)$. The symmetrized discord
\be
D_3^\mathrm{{sym}}:=I(\rho_{AB})-J_3^{\mathrm{sym}}(\rho_{AB}),
\ee
is called the measurement induced disturbance and serves as another  upper
bound on the discord.

\section{Maxwell's goblins}

Maxwell's demon \cite{max} is a ``being whose facilities are so sharpened" as to enable him to challenge the 2\textsuperscript{nd} law of thermodynamics. Modern exorcism mostly focuses on his information-processing ability, with information erasure cost balancing the books and keeping the 2\textsuperscript{nd} law intact. Quantum logic and quantum correlations introduce new subtleties into this discussion \cite{md2}.

A typical setting is provided by a (quantum) Szilard's model \cite{szi}, in whose original form the demon operates a heat engine with one-particle working fluid. For our purposes  it is enough to consider only the work-extracting phase of the cycle and ignore the the resetting of the demon's memory.  The optimal work extracted from a system of a dimension $d$ in a known state $\rho$  at the temperature $T$ is on average
\be
W^+=kT(\log d-S(\rho))=kTK(\rho), \label{demon}
\ee
where $k$ is Boltzman's constant adjusted to the base of the logarithm.

For a bipartite state $\rho_{AB}$ the benchmark performance $W^+(\rho_{AB})$ is achieved by a fully quantum (non-local) demon Charlie that can perform arbitrary  quantum operations on the system. We compare his performance with actions of two local goblins of lesser power. Alice and Bob are local goblins that can perform only local operations on their subsystems. They may have only partial information about the state $\rho_{AB}$  and may not be able to communicate freely.

The work that is extracted by Alice and Bob that are aware of their respective states $\rho_{A,B}$ but are not allowed to communicated is
\be
W_L=kT(\log d_Ad_B-S(\rho_A)-S(\rho_B)),
\ee
so the difference of the extracted work by a global demon and local non-communicating goblins is given by the mutual information,
\be
\Delta_L W:=W^+-W_L=kT I(\rho_{AB}).
\ee

A much more interesting scenario that actually motivated the introduction of $D_2$ was proposed in \cite{z03}. In this setting both Alice and Bob know the state $\rho_{AB}$ and Alice can communicate to Bob the results of her measurement. She chooses her measurement $\Pi$ in such a way as to maximize the extracted work
\begin{align}
W_2 & =[\log d_A -H(A_\rho^\Pi)]+ [\log d_B- S(\rho_B|\Pi^A) \nonumber \\
& =\log{d_Ad_B}-S(\rho_{AB}^{\Pi^A}),
\end{align}
through steering of Bob's state to $\rho_{B|\Pi_a}$ which on average make up for a higher entropy of  $\rho^{\Pi^A}_A=\sum_a\Pi_a\rho_A\Pi_a$. Hence the minimal difference between  the work extracted by the goblins and  the work extracted by the  demon is given by
\be
\Delta_2W=kTD_2^A(\rho_{AB}).
\ee
In this setting Alice and Bob are essentially performing the purification protocol from Sec IIC.

An operational meaning of $D_3$ is clarified in the setting where Alice is still able to report her results to Bob, but has less knowledge than in the original example. Namely, Alice knows only $\rho_A$, while Bob is aware of the entire state $\rho_{AB}$. In this case the best Alice can do is to perform the measurement in the eigenbasis of $\rho$, and tell her result to Bob. Then on average the gain is
\be
W_3=kT(\log d_A-S(\rho_A))+kT(\log d_B-S(\rho_B|\Pi^A_*)),
\ee
so the difference in the extracted work is now
\be
\Delta_3W=kTD_3^A(\rho_{AB}).
\ee

\section{Local distinguishability and discord}

Since zero discord is thought to represent the absence of classical correlations, it is interesting to investigate the following question. Consider a set of pure orthogonal bipartite  states, each of which may have a different prior probability, with the ensemble density matrix $\rho_{AB}$.  Does the  value of $D(\rho_{AB})$ tell us something about the ability to perfectly distinguish these states by local operations and classical communication (LOCC)?

As exhibited Table I  , there is no relation  between $D(\rho_{AB})$ and  local distinguishability.  First, it was shown in \cite{tea99} that a certain set of nine $3\times 3$ product orthogonal states cannot be perfectly distinguished by LOCC. These ``sausage"  (or  ``four and a half tatami teahouse") states $|\psi_1\9,\ldots|\psi_9\9$ are
\begin{align}
|1\9\otimes |1\9, \quad |0\9\otimes|0\pm 1\9/\sqrt{2},\quad |2\9\otimes|1\pm2\9/\sqrt{2}, \nonumber\\
|1\pm2\9\otimes|0\9/\sqrt{2}, \quad |0\pm 1\9|2\9/\sqrt{2},
\end{align}
Their equal mixture is the maximally mixed state $\rho_{AB}=\1/9$ that obviously has a zero discord, $D^A(\rho_{AB})=D^B(\rho_{AB})=0$.

Second, it was shown \cite{wv00} that any two orthogonal (entangled or not) states can be perfectly distinguished by LOCC. Consider a mixture
of the Bell states $|\Psi^\pm\9$,
\be
\rho_{AB}=a |\Psi^+\9\6\Psi^+| +(1-a) |\Psi^-\9\6\Psi^-|,  \qquad 0<a<1.
\ee
The discord of such a state can be calculated analytically. It equals to
\be
D_{1,2}^{A,B}(a)=a\log_2a-(1-a)\log_2a+1,
\ee
which vanishes only for the equal mixture $a=\half$.

Two other cases are easy. A pair of bi-orthogonal product states, such as $|0\9|0\9$ and $|1\9|1\9$ results in a zero discord.
Mixing nine teahouse states with different weights may result in a non-zero discord. For example, giving  $|\psi_9\9$ and $|\psi_7\9$ weights which are twice as high as  the rest of the states results in a mixture $\rho_{AB}$ for which $[\rho_A\otimes\1_B,\rho_{AB}]\neq 0$, thus implying a non-zero discord.

\begin{table}[htbp]
   \centering
    \caption{Local measurability vs. discord}     \label{Discordtab}
    \begin{tabular}{|ll|c|}
          \toprule
      \multicolumn{3}{c}{} \\
      \cmidrule(r){1-2}
      \hline
     ~~~~~~~~~~~~~~~~States    & Discord & Locally \\
     & & measurable\\
      \hline
     9 teahouse states, equal weights   & $D^A=D^B=0$ & no  \\
          2 product bi-orthogonal states   & $D^A=D^B=0$     &  yes \\
         2  entangled orthogonal  states  & $D_1^A>0$  & yes  \\
     9 teahouse states, unequal weights    & $D_1^A>0$  & no \\
     \hline
   \end{tabular}
   \label{tab:booktabs}
\end{table}

\section{Summary and outlook}
There are at least three useful one-way measures of  the quantumness of states, namely the three discords. They vanish simultaneously, and it is easy to check if this is the case, using Properties 1 and 2. The discord measures $D_2$ and $D_3$  have a natural physical interpretation in terms of the work extracted by a pair of Maxwell's demons that operate on a bipartite system under different sets of restrictions.
On the other hand, depending on the imposed restrictions, it is either $D_1$ or $D_2$ that can serve as a measure of quantum deficit in the state.

Despite its intuitive appeal quantum  discord $D(\rho_{AB})$ is not an indicator of wether the  ensemble of states $\rho_{AB}$ is made up of states which are distinguishable by LOCC. Zero quantum discord allows for a completely-positive dynamics of an open system even if there are initial correlations with the environment \cite{discp}. In a forthcoming paper \cite{gf} it will be shown that it is not enough for practical quantum tomography.

We discussed the discord in the states on finite-dimensional spaces. One obvious difficulty in the generalization to the continuous variables is in the infinite-dimensional optimization that is involved in the definition of $D_1$ and $D_2$. It is possible that the measure $D_3$ is more suitable in the latter case.

\bigskip

\acknowledgments
We thank Daniel Cavalcanti, Animesh Datta, Mile Gu, Micha{\l},  Pawe{\l},  and Ryszard Horodecki, Jason Twamley,  Vlatko Vedral, Wojciech Zurek, and Karol \.{Z}{y}czkowski for discussions and helpful comments.

\end{document}